\documentclass[conference]{IEEEtran}
\IEEEoverridecommandlockouts
\usepackage{cite}
\usepackage{amsmath,amssymb,amsfonts,amsthm}
\usepackage[linesnumbered,ruled,vlined]{algorithm2e}
\usepackage{algpseudocode}
\usepackage{graphicx}
\usepackage{textcomp}
\usepackage[table]{xcolor}
\usepackage{rotating}
\usepackage{subcaption}
\usepackage{multirow}
\usepackage{color,soul}
\usepackage{url}
\usepackage{balance}
\usepackage{pifont}
\usepackage[font=small]{caption}
\usepackage[inline]{enumitem}
\usepackage{amsmath}
\usepackage{amssymb}
\usepackage{multirow}
\usepackage{amsthm}
\usepackage{braket}

\def\BibTeX{{\rm B\kern-.05em{\sc i\kern-.025em b}\kern-.08em
    T\kern-.1667em\lower.7ex\hbox{E}\kern-.125emX}}

\usepackage{todonotes}
\theoremstyle{remark}

\begin{document}

\title{Moving Target Defense for Service-oriented Mission-critical Networks
}
\author{\IEEEauthorblockN{Do\u{g}analp Ergen\c{c}}
\IEEEauthorblockA{University Hamburg}
ergenc@informatik.uni-hamburg.de
}

\author{
	\IEEEauthorblockN{Do\u{g}analp Ergen\c{c}, Florian Schneider, Peter Kling, Mathias Fischer}
	\IEEEauthorblockA{\textit{Universit{\"a}t Hamburg}, DE}
	name.surname@uni-hamburg.de
}
\maketitle

\begin{abstract}
Modern mission-critical systems (MCS) are increasingly softwarized and interconnected. As a result, their complexity increased, and so their vulnerability against cyber-attacks. The current adoption of virtualization and service-oriented architectures (SOA) in MCSs provides additional flexibility that can be leveraged to withstand and mitigate attacks, e.g., by moving critical services or data flows. This enables the deployment of strategies for moving target defense (MTD), which allows stripping attackers of their asymmetric advantage from the long reconnaissance of MCSs. However, it is challenging to design MTD strategies, given the diverse threat landscape, resource limitations, and potential degradation in service availability. In this paper, we combine two optimization models to explore feasible service configurations for SOA-based systems and to derive subsequent MTD actions with their time schedule based on an attacker-defender game. Our results indicate that even for challenging and diverse attack scenarios, our models can defend the system by up to 90\% of the system operation time with a limited MTD defender budget.
\end{abstract}


\section{Introduction}
Modern mission-critical systems~(MCSs), like smart cars and avionics, consist of interconnected services that carry out collaborative tasks. This results in additional complexity and thus, a broader surface for cyber-attacks.
To cope with the additional complexity, service-oriented architectures~(SOA) and virtualization are increasingly adopted in different mission-critical domains~\cite{Ergenc2020, Villaneueva2021, Cucinotta2009}.
SOA can accommodate the system design by enabling flexible and isolated service deployment on virtualized hardware. 
Such flexibility also enables a reconfiguration of systems to handle failures and to withstand and recover from cyber-attacks. 

From a security perspective, attackers have an asymmetric advantage against traditional MCSs since they can conduct a long reconnaissance before they carry out their attacks~\cite{Chen2011}. Besides, an attacker can remain in stealth for months to make the highest impact even after infiltrating a system~\cite{fireeye2020}. Here, the longer the system remains in its static configuration, the higher the probability of a successful attack is. Defenders, however, have only a limited time to detect and mitigate it. 
Moving target defense~(MTD) can balance this asymmetry by reconfiguring critical assets~\cite{Sengupta2020}, e.g., shuffling IP addresses or changing the allocation of critical services. It renders the attacker's knowledge about the system obsolete and thus impedes attacks. 
SOA and virtualization ease the development of MTD strategies as they enable the migration and replacement of services and reconfiguration of their inter-communication.

However, MTD via service reconfiguration requires additional spare resources and induces reconfiguration costs for increased delay and packet loss. 
Furthermore, without a precise understanding of potential attacks and failures, an MTD strategy causes too frequent or ineffective reconfigurations~\cite{Zhang2019, Thompson2014, Thompson2016}.
Therefore, we need an effective MTD strategy that determines \emph{which} services must be changed, \emph{how} they are changed (e.g., migrate or re-instantiate), and \emph{when} they are changed.
To address those questions, various attacker-defender games have been proposed in the context of game theory, e.g., FlipIT~\cite{Van2013} or the probabilistic learning attacker and dynamic defender~(PLADD) model~\cite{Jones2015}. 
Although they have already derived asymptotical bounds for optimal MTD strategies, they do not provide concrete steps to reconfigure systems.
Moreover, these models do not include network design constraints for resource management and quality of service, which is especially important for MCSs. 

This paper proposes an optimization framework to determine subsequent service configurations within optimal MTD strategies based on an attacker-defender game. Accordingly, our contributions are:
\begin{itemize}[leftmargin=*]
\item We repurpose our linear programming model for joint service allocation and routing~(JSAR)~\cite{Ergenc2020, Ergenc2021} to identify a set of feasible service configurations satisfying resource and QoS requirements of SOA-based MCSs. 
\item We formulate a novel optimization model, PLADD-scheduling~(PLSCH) based-on the PLADD game~\cite{Jones2015}, to find optimal MTD schedules against various attacks. 
\item We develop a composite model, PLSCH-MTD, to deploy the resulting configurations of JSAR for each MTD action over the time-schedule provided by PLSCH.
\item We create several attack scenarios reflecting the time characteristics of recent security incidents in MCSs to evaluate PLSCH-MTD.
\end{itemize}

In the rest of the paper, Section~\ref{sec:background} introduces the preliminaries for the considered attack-defender game. Section~\ref{sec:related} presents related work on SOA design and MTD. Section~\ref{sec:design} introduces our optimization models JSAR, PLSCH, and PLSCH-MTD.
Section~\ref{sec:attacker} describes the attack scenarios that are used to evaluate PLSCH-MTD.
Section~\ref{sec:eval} presents the evaluation results, and Section~\ref{sec:conclusion} concludes the paper.

\section{Background} \label{sec:background}
In this section, we describe two essential concepts of this study: The probabilistic learning attacker and dynamic defender~(PLADD) model and the PLADD-scheduling problem. We also note our assumptions and modifications that make the formulation of these models more convenient for this study. 

\subsection{Probabilistic Learning Attacker and Dynamic Defender}
\begin{figure}[h!]
\center
\includegraphics[scale=0.5]{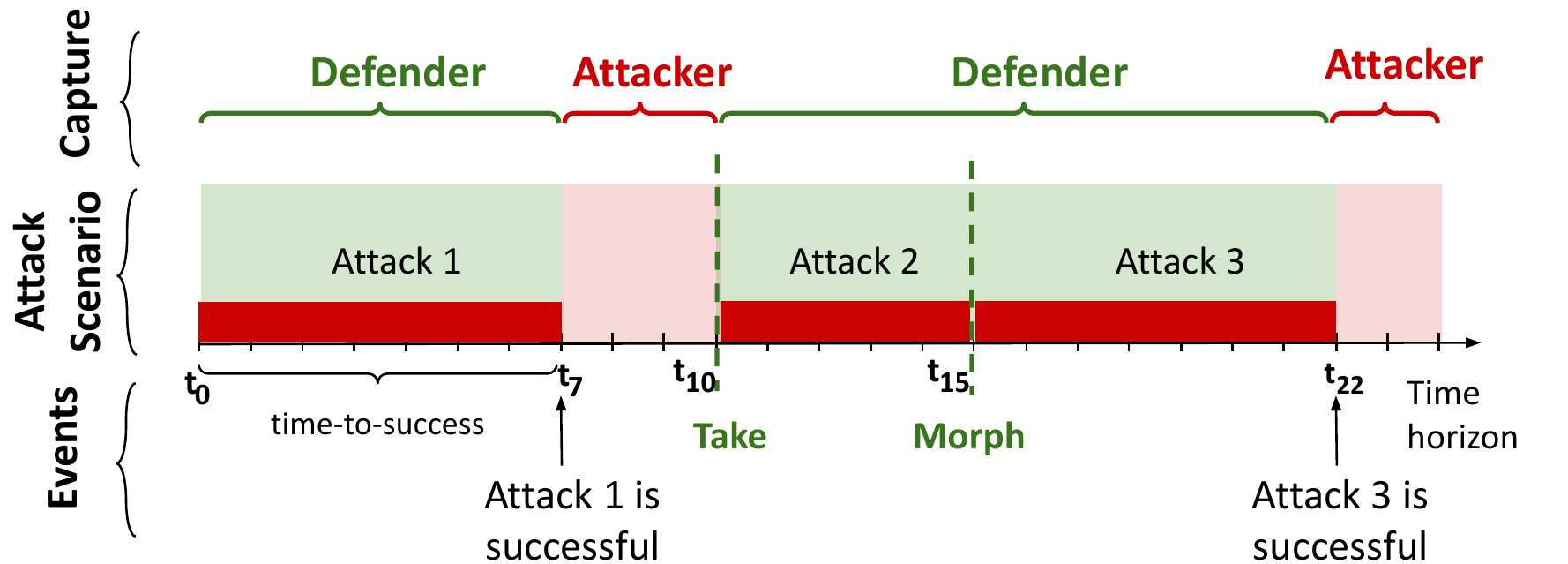}
\caption{An example of the PLADD game.}
\label{fig:pladd_disc}
\end{figure}

PLADD introduces an attacker-defender game that involves (i) an attacker with learning capabilities and (ii) a defender with various actions competing to gain control of the system within a given \textit{time horizon} that represents a certain frame of the system's operation time~\cite{Jones2015}.
Fig.~\ref{fig:pladd_disc} shows its fundamentals.
An attacker can conduct successive attacks (red blocks) that each takes a certain time to be completed, i.e., having \textit{time-to-success}. As a result of a successful attack, the attacker captures the resources (indicated by the light red background, e.g.~from $t_7$ to $t_{10}$).
When an attacker completes an attack, it might \textit{learn} about the system, and its subsequent attack takes less time accordingly, e.g., attack~2 is shorter than attack~1. 

The role of a defender is to conduct certain actions (vertical dashed lines) to prevent an attacker from completing its attack. A \textit{take} action usually represents an instant intervention, e.g., resetting a service instance, while a \textit{morph} action refers to more substantial system changes, e.g., migrating multiple services over the system nodes with diverse configurations.
After the defender morphs the system, the attacker loses her knowledge obtained after successful attacks and thus should spend a longer time for its upcoming attacks (e.g., attack 3 in Fig.~\ref{fig:pladd_disc} is longer than attack 2).
Similar to the \textit{take} action, the defender captures the resources back after \textit{morph}. 
   
Both an attacker and a defender have limited \textit{budget}.
An attacker can have only limited attacking opportunities, and lengthy attacks require more effort.
A defender cannot reconfigure the system too often, and the cost of a reconfiguration is usually proportional to the changes in the system, as they usually cause service interruptions.
Therefore, the defender should conduct its actions within an effective time-schedule against potential attacks within its budget.

\emph{Eventually, in the PLADD model, the goal of the attacker is to complete a sequence of attacks and gain control over the system after each successful attack. The defender aims to develop a strategy that determines (i) the type of defensive action to prevent an attack and (ii) a schedule to conduct a sequence of actions against particular attack scenarios within its limited budget.} 

We mainly focus on PLADD as it can model different types of MTD actions and their effective scheduling to minimize the attacker's advantage. Furthermore, it can capture the time characteristics of several attacks, which can vary from relatively fast reconnaissance attempts to long-term advanced persistence threats~(APTs) in MCSs. 
In our formulation, we assume that the time-to-success of an attack is independent of the previous successful attacks, i.e., the attacker does not learn. It enables us to develop defensive strategies against potential attack scenarios, whose characteristics can be modeled in advance. As a result, we consider a single type of defensive action (\textit{take} or \textit{morph}), referred to as the \textit{infinite} model in~\cite{Jones2015}. 
This action corresponds to \textit{morph} in the original PLADD model regarding its impact since the service migrations over the system lead to a significant reconfiguration.
Lastly, we have not limited the attacker to a certain budget and assume that it can conduct attacks whenever the defender regains control over the system. 

\subsection{PLADD-Scheduling}\label{sec:plsch}

PLADD-Scheduling~(PLSCH) leverages the PLADD game to provide an exact schedule for the defensive strategy, e.g., when to conduct \textit{take} or \textit{morph} actions.
Originally proposed in~\cite{osti_1511904}, it formulates the attacker-defender game as a combinatorial job assignment problem. Here, we first describe the job assignment problem and then explain how it corresponds to the original PLADD model. 
It considers a system of $m \in \mathbb{N}$ machines over a time horizon of $T \in \mathbb{R}^+$ time units.
Each machine $m$ comes with a job sequence $J_m = (d_{m1}, d_{m2}, \dots)$ of at most $n+1$ jobs which it must process.
The \emph{duration} $d_{mj} > 0$ of the $j$th job on machine $m$ specifies how long it takes machine $m$ to process its $j$th job. 
In order to start processing the $j$th job of $J_m$, machine $m$ must have finished the first $j-1$ jobs. A job on any machine can only start after a \emph{starting action}, which affects all machines simultaneously, is taken. However, the number of those actions is limited and thus, they should be scheduled effectively to initiate several jobs across multiple machines. The time between the end of a job and the beginning of a subsequent job, i.e., if the jobs cannot be scheduled adjacently, is idle machine time.

To see the connection to the PLADD game, we interpret each machine as one of $m$ possible, equally likely attack scenarios. 
A job $j$ corresponds to an individual attack within an attack scenario, and its duration is the time-to-success value for the respective attack. A finished job means that the attacker captured the resources.
The starting action for the jobs corresponds to the instantaneous time that the defender retakes control of a potentially compromised system, i.e., a \textit{take} action in the PLADD model. The limitation on the number of starting actions represents a limited defender budget. Note that the attacker continuously conducts attacks right after each defender action competing for the system resources. 

With this interpretation, the goal in PLSCH becomes to schedule the jobs such that the total idle time over all machines is minimized. This time also corresponds to the duration when the resources are under the attacker's control.
Note that minimizing the idle time is equivalent to maximizing the total time any machine is active~(not idle).

\begin{figure}[h!]
\center
\includegraphics[scale=0.54]{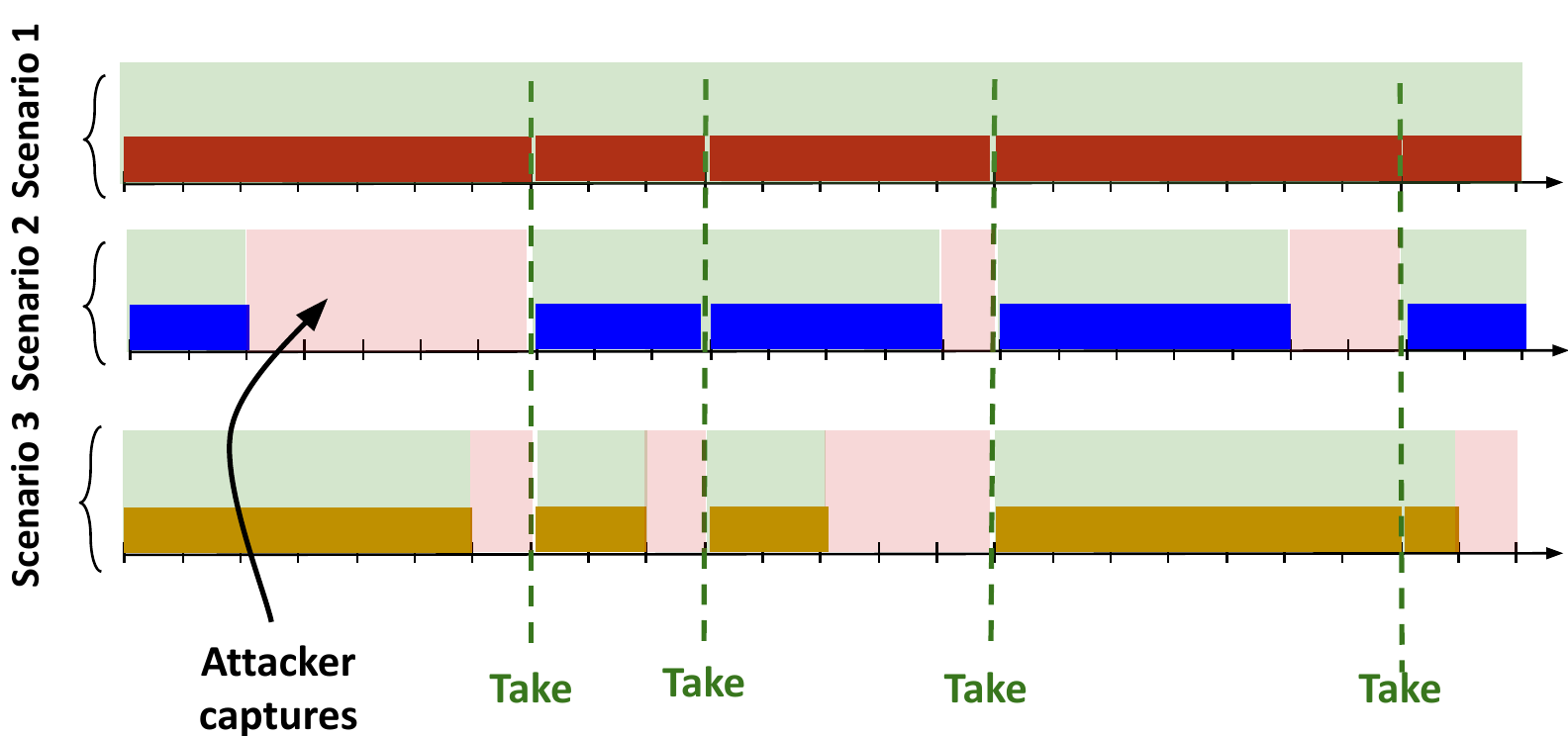}
\caption{A single defensive schedule for multiple attack scenarios.}
\label{fig:pladd_multimachine}
\end{figure}

In PLSCH, a single action determines the job assignment on multiple machines. That is, the jobs on all machines initiate simultaneously according to a single schedule of starting actions.
The reason lies in the formulation of the PLADD model: The defender cannot know the actual attack scenario and, thus, must develop the most effective strategy to defend against all likely attack scenarios. Accordingly, all the jobs are generated in advance as input to the model, reflecting the potential attack scenarios against the target system. Fig.~\ref{fig:pladd_multimachine} illustrates the difficulty of scheduling jobs over multiple machines, which corresponds to protecting the system against various attack scenarios simultaneously. While there is no idle time on the first machine, i.e., complete protection against the first attack scenario, the same schedule results in more idle times on the other machines, i.e., resulting in the attacker's success.

\section{Related Work} \label{sec:related} 
In this section, we present the state-of-the-art on (i) SOA-based network design, (ii) moving target defense, and (iii) game-theoretical approaches for network security. 

\textbf{Service Distribution and Network Design:}
In SOA-based mission-critical networks, the critical services and flows are the assets to be protected. Therefore, a reconfiguration in the context of MTD involves a service allocation and flow assignment problem. 
A proper service allocation~\cite{Li2016, Yi2018} is important to, for instance, minimize operational costs \cite{Addis2015} and physical resource fragmentation~\cite{Bari2016} for the providers, and maximize the service quality \cite{Addis2015} and responsiveness \cite{Liu2017} for the user experience.
It usually requires an accurate resource orchestration regarding where, when, and how many service instances are deployed~\cite{ChenniKumaran2018, Pandey2017}. Besides, the dependencies of services on each other~\cite{Espling2016}, service migrations \cite{Breitgand2011}, load-balancing~\cite{Pu2018}, task scheduling \cite{Gawali2018}, and  power-awareness~\cite{Varasteh2021} are some of the design constraints that are addressed in the literature. 
Other studies address the service allocation and routing problem jointly to deploy the services on the paths aiming for optimal resource utilization~\cite{Luizelli2015, Lee2015}.
Recent studies include the service protection and availability issues as well~\cite{Ergenc2020, Ergenc2021, Askari2021}.

In this work, we use our previous service allocation and flow assignment scheme~\cite{Ergenc2020} as it directly reflects the SOA requirements of the MCSs. Moreover, it offers configurability with dynamic services and flows, which gives a large reconfiguration space for potential MTD strategies. 

\textbf{Moving Target Defense:}
MTD is a well-studied field that enables the development of defensive strategies by moving the critical assets in a system.
The authors of \cite{okhravi2012creating} implement multiple diverse platforms with different software packages, operating systems, and processor architectures. The system functions are then moved among such platforms keeping the state information. Similarly, in~\cite{Huang2011}, a pool of diverse virtual systems is orchestrated by a controller to fluctuate the attack surface by switching on and off the redundant resources on different components. In~\cite{Crouse2011}, it is argued that any configuration parameter may impact the overall security. They propose a genetic algorithm to find the best suitable (re)configuration to minimize the chance of a successful attack. The authors of \cite{al2011toward} focus on mutating the network configuration, e.g., IP addresses, ports, and destination addresses. In \cite{Thompson2014} and \cite{Thompson2016}, they circulate the virtual machines with different operating systems as well as change network addressing schemes to prevent both OS- and network-targeted persistent attacks.

In contrast to the related work, we consider the services and flows as our critical assets in the SOA context. Although several studies merely focus on migrating virtual instances, we propose an optimization framework to find feasible and timely system-wide reconfiguration.  

\textbf{Game theory:}
Game theory offers solid analytical tools to develop attacker and defender interactions to develop effective defensive strategies~\cite{Alese2014, Liu2021}.
The same authors of PLADD extend their evaluation with further insights in~\cite{Jones2017}. According to the practical implications of the study, it is always possible to push a rational attacker out of the PLADD game even though it might not be cost-optimal for the defender. 
PLADD has also been considered for the security modeling of various networking areas. In~\cite{Chen2021}, the authors utilize PLADD to defend power grid infrastructure. They analyze the optimal schedule to reset access controls of the system to minimize the probability of a successful attack. 
In~\cite{Gao2020}, the authors focus on the multi-attacker and defender games for massive machine-type communications (mMTC) in 5G. They formulate a non-zero-sum differential game with attack and defense alliances and propose an optimal defensive strategy algorithm. 
In \cite{Xu2017}, the authors address APTs toward cloud systems. In this game model, two parties compete to set their attack and scan intervals based on their subjective decisions. 
In \cite{Li2020} and \cite{Sengupta17}, the authors formulate spatio-temporal Stackelberg games to find optimal configurations for web applications over time.
 
In comparison to the related work, we model various attack scenarios in terms of their time characteristics against MCSs rather than focusing on smaller-scale web applications. We also evaluate different types of attacks beyond specific vulnerabilities of their target applications. In addition, we address the complex interdependencies of connected services regarding their resource consumption and QoS in SOA-based MCSs. 

\section{PLADD-Scheduling MTD (PLSCH-MTD) Optimization Model} \label{sec:design}

\subsection{Solution Overview}

\begin{figure}[ht!]
\center
\includegraphics[scale=0.32]{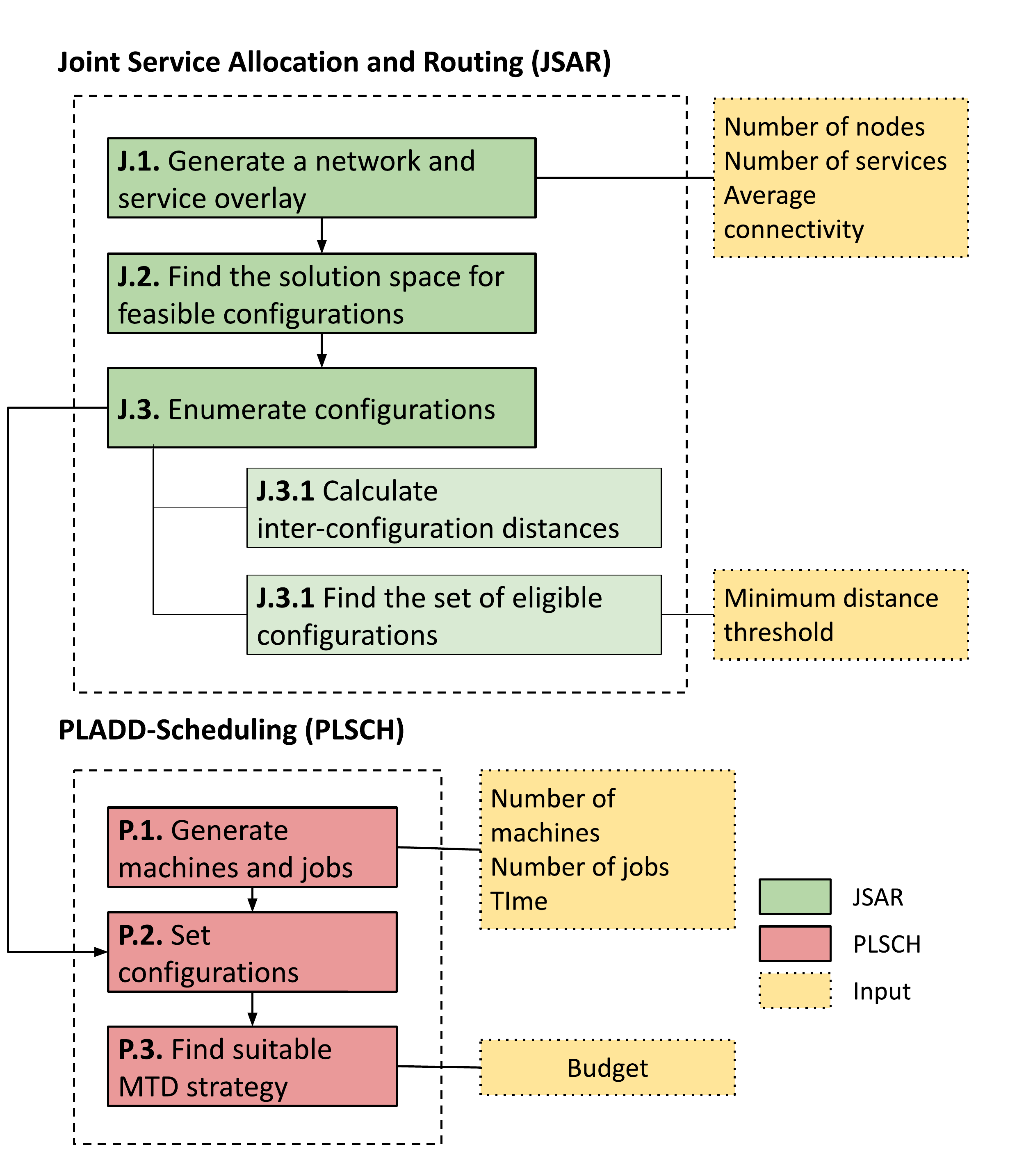}
\caption{The overall optimization framework: PLSCH-MTD.}
\label{fig:overview}
\end{figure}

Fig.~\ref{fig:overview} shows the steps of the overall model, PLSCH-MTD, which consists of two optimization processes: (i) Joint Service Allocation and Routing~(JSAR) and (ii) PLADD-Scheduling~(PLSCH). While JSAR provides the possible configurations to be used within MTD actions (green blocks), PLSCH finds a schedule for the defender's actions by changing respective configurations to defend against the considered attack scenarios (red blocks). In the rest of this section, we discuss, first, the PLSCH process for scheduling and then present the integration of MTD to the given model accordingly.

\subsection{PLADD-Scheduling (PLSCH) Model} \label{sec:plsch-formulation}

In this section, we formulate the PLSCH problem described in Section~\ref{sec:plsch} as an integer linear program (ILP). The idea of such combinatorial model is introduced in~\cite{osti_1511904} and we modify and extend the model by implementing time-to-success requirements and with further constraints. Table~\ref{tab:plsch} shows all related variables and parameters.

\begin{table}[ht!]
\caption{Variables and parameters of PLSCH.}
\label{tab:plsch}
\centering
\begin{tabular}{|c||c|c|l|} 
 \hline
 \textbf{Type} & \textbf{Symbol} & \textbf{Set} & \textbf{Definition} \\ [0.5ex] 
 \hline\hline
 \multirow{3}{*}{Base} & $m$ & $M$ & A machine \\ \cline{2-4}
 & $j, k$ & $J_m$  & A job to be scheduled on machine $m$ \\ \cline{2-4}
 & $t, u$ & $Z^{*}$ & Discrete time instance \\ \hline
 \multirow{2}{*}{Constant} & $d_{mj}$ & $Z^{*}$    & Duration of job $j$ on machine $m$\\ \cline{2-4}
 & $\beta$ & $Z^{*}$ & Defender budget \\ \hline 
  \multirow{2}{*}{Variable} & $x_{t}$ & $Z^{*}$ & Decides if an action taken at $t$\\ \cline{2-4}
  & $y_{mjt}$ & $Z^{*}$ & Decides if $j$ scheduled on $m$ at $t$ \\ \hline
 \end{tabular}
\end{table}

As explained in Section~\ref{sec:plsch}, we represent each attack scenario as a \textit{machine} and individual attacks in each scenario as \textit{jobs} in the PLSCH model.
The PLSCH takes (i) a number of machines with different sequences of jobs and (ii) a fixed action budget as input. It provides a schedule for the multiple-machine job assignment problem that corresponds an MTD schedule for the defender. Accordingly, there are two optimization variables, $x_t$ and $y_{mjt}$. $x_t$ is a binary decision variable that represents if any starting action is schedule at the time instance $t < T$ to initiate a job, where $T$ is the system's operational time, i.e., time horizon. $y_{mjt}$ is the other binary variable to decide if a job $j \in J_m$ of machine $m \in M$ is scheduled to start at the time instance $t$. Note that the number of starting actions are limited by the action budget, which corresponds to the defender budget in PLADD. All other constraints are given as follows. 

\begin{align}
& \sum_{t=0}^{T}{y_{mjt}} \leq 1 & \forall m \in M, \forall j \in J_m \label{const:uniqueness} 
\end{align}
Constraint~\ref{const:uniqueness} ensures that $j$ can be scheduled only once on $m$.
\begin{align}
& \sum_{t=0}^{T}{y_{mjt}} - \sum_{t=0}^{T}{y_{mkt}} \geq 0 & \forall m \in M, \forall j,k \in J_{m}, k = j +1 \label{const:order} 
\end{align}
Constraint~\ref{const:order} ensures that a job $k$ can take place on machine $m$ only if its predecessor job $j$, i.e., $k = j + 1$, is scheduled on $m$ at a time instance $t$. It implies that a job $k$ cannot be scheduled on a machine $m$ before all other jobs $j$ in the job set $J_m$ s.t., $j < k$ are placed. Accordingly, all attacks in each attack scenario are defended in the given order. 
\begin{align}
(d_{mj} + t)y_{mjt} + (T - u)y_{mku} \leq T \qquad &  &  \notag  \\
  \qquad \qquad  \forall m \in M, \forall j,k \in J_m, k = j+1, \forall t,u \leq T  & &   \label{const:noninterference} 
\end{align}
Constraint~\ref{const:noninterference} ensures that two consecutive jobs $j$ and $k$ s.t. $k = j + 1$ in a single attack scenario cannot overlap as the successor job $k$ restricted to start after the whole duration of $j$, $d_{mj}$, s.t. $t + d_{mj} \leq u$, where $t$ and $u$ are the starting times of $j$ and $k$, respectively. Besides, the finishing time of $j$ is constrained by the total operating time of the system, $T$, in case there is not successor job scheduled. Note that non-overlapping jobs in a machine in PLSCH formulation could imply that there cannot be concurrent attacks in an attack scenario in PLADD game. However, the PLSCH handles that by introducing multiple machines so that the resulting strategy can defend against multiple attack scenarios simultaneously. 

\begin{align}
& \sum_{j \in J_m}^{}{y_{mjt}} \leq x_{t} & \forall m \in M, \forall t \leq T \label{const:take} \\
& \sum_{m \in M}^{}{\sum_{j \in J_m}^{}{y_{mjt}}} \geq x_{t} & \forall t \leq T \label{const:nonempty} 
\end{align}
Constraints~\ref{const:take} and \ref{const:nonempty} represent the dependencies between two decision variables. Constraint~\ref{const:take} ensures that (i) no job $j$ can be scheduled at $t$ unless there is a starting action s.t., $x_t = 1$ and (ii) at most one job can be placed on machine $m$ at a given time instance $t$. Complementarily, constraint~\ref{const:nonempty} implies that there should be at least a job scheduled in one of the machines if a starting action takes place at the given time. Those constraints also model the dynamics of the PLADD game, s.t., an attacker is expected to conduct a new attack right after the defender takes an action and regains the control.  

\begin{align}
& \sum_{t=0}^{T}{x_t} \leq \beta & \label{const:budget} 
\end{align}
Lastly, constraint~\ref{const:budget} limits the number of starting actions by $\beta$. In PLADD, it corresponds to the limited defender budget in terms of the number of MTD actions.

The objective function~\ref{func:maxjob} maximizes the occupation of a machine with respective jobs. This corresponds to the time spent by the attacker to conduct attacks when the defender holds control of the system. It also implies the minimization of the idle time of all the machines, i.e., decrease the time when the attacker captures the system~\cite{osti_1511904}. Therefore, it eventually aims to protect the system from being occupied by the attacker considering all (given) potential attack scenarios. 
\begin{align}
\max & \sum_{m \in M}^{}{\sum_{j \in J_m}^{}{\sum_{t=0}^{T}{y_{mjt}d_{mj}}}}& \label{func:maxjob} 
\end{align}


\subsection{JSAR: Network Configuration Model} \label{sec:jsar}
Besides scheduling MTD actions, the defender must decide which configuration to apply to take adequate measures. A configuration consists of (i) allocating mixed-criticality services over virtualized MCS nodes and (ii) establishing their intercommunication within limited system resources. This decision is highly dependent on the structure of the network. 
In~\cite{Ergenc2020}, we proposed JSAR as an optimization model for the design of mission-critical networks according to the given definition.
Here, we utilize the model to generate a feasible solution space, e.g., a set of sub-optimal configurations, that can be used by the defender to change the deployment of the network. We present the details of JSAR in this section. 

The JSAR takes (i) a network of nodes with different processing capacities and connected via links with limited bandwidth and (ii) a service overlay with inter-connected services with certain QoS demands. 
$z_{dp}$ and $q_{sv}$ are two binary decision variables that represent if demand $d$ is assigned to path $p$ and if service $s$ is deployed on node $v$, respectively. 
The objective function~(\ref{const:bootstrapping_obj}) minimizes the length of selected paths, where $|p|$ represents the path length. 
Minimizing the total path length can be considered as both performance and cost optimization by establishing low-latency communications, i.e., here with fewer hops, and decreasing the number of occupied links. Depending on the various goals of the defender as a network designer, the objective function can be easily adapted. 

Constraint~(\ref{const:node_capacity}) and (\ref{const:node_capability}) ensure that $v$ has sufficient resources to host $s$ and $s$ is deployed on exactly one node that is capable to host $s$ (e.g., equipped with the required hardware). Constraint~(\ref{const:path_service_assignment}) restricts that $d$ can be deployed on $p$ only if the required services $s$ and $t$ are deployed on the source and destination nodes of path $p$, which are $u$ and $v$. This quadratic constraint is linearized using McCormick envelopes~\cite{Mccormick1976}. Constraint (\ref{const:link_capacity}) ensures that each link $e$ of $p$ has sufficient resources to carry the traffic of $d$ if it is assigned to $p$. While constraint (\ref{const:latency}) ensures that $p$ is selected to satisfy the maximum tolerable latency for $d$, constraint (\ref{const:path_assignment}) guarantees that $d$ is assigned exactly to one path. 
\begin{align}
\min & \sum_{d \in D}^{}{\sum_{p \in P}^{}{z_{dp} |p|}} &  \label{const:bootstrapping_obj} \\
& \sum_{s \in S}^{}{q_{sv}\tau_s} \leq r_v & \forall v \in V \label{const:node_capacity}\\
& \sum_{v \in V}^{}{o_{sv}q_{sv}} = 1 & \forall s \in S \label{const:node_capability}\\
& z_{dp} \leq q_{sv}q_{tu} + q_{tv}q_{su} & \forall d \in D, \forall u,v \in V, \notag  \label{const:path_service_assignment} \\
&  & \forall p \in P_{uv}, (s,t) \in d  \\
& \sum_{d \in D}^{}{\sum_{\substack{p \in P, \\ e \in p}}^{}{z_{dp}h_d}} \leq c_e & \forall e \in E  \label{const:link_capacity}\\
& \sum_{e \in p}^{}{z_{dp} l_e^*} \leq l_d & \forall d \in D, \forall p \in P    \label{const:latency}\\
& \sum_{p \in P}^{}{z_{dp}} \geq 1 & \forall d \in D    \label{const:path_assignment}
\end{align}

\subsection{PLSCH-MTD: Integrating PLSCH and JSAR}

In this section, we propose the integrated model, PLSCH-MTD, to make use of a set of feasible network configurations according to their eligibility together with PLADD schedules. We extend PLSCH to set suitable configurations obtained via JSAR for each scheduled action of the defender. An action represents the replacement of services and re-routing, consuming the limited budget of the defender. At the same time, we enforce a minimum amount of changes between successive configurations. Note that changing the whole configuration may force the attacker to perform a complicated attack once more, but it comes with a certain cost to re-design the network. Eventually, the distance between two configurations deduces a trade-off between reconfiguration overhead and defensive capabilities, e.g., creating a degree of obscurity. 

Having a large set of configurations, a defender should decide which configurations can be set after a particular configuration, e.g., which are eligible to be the next configuration. To quantify the eligibility, we propose the following metric, \textit{distance} between two configurations. It is calculated between two configurations $c$ and $e$ as
\begin{align}
|c - e| = \frac{\sum_{s \in S}^{}{\sum_{v \in V}^{}{|q_{sv}^c - q_{sv}^e|}} +  \sum_{d \in D}^{}{\sum_{p \in P}^{}{|z_{dp}^c - z_{dp}^e|}}}{|S| + |D|} \label{eq:distance} 
\end{align}
where $q_{sv}^c$ and $z_{dp}^c$ represent the service deployment and demand assignment variables (cf. JSAR) for the configuration $c$, respectively. The distance between two configurations $c$ and $e$ is proportional to (i) the number of service migrations, i.e., services migrated to different nodes than the previous configurations, and (ii) reroutings, i.e., traffic streams moved to different paths. Eq.~\ref{eq:distance} can also be used to calculate the migration overhead that may cause a certain delay and configuration effort for each reconfigured component. We evaluate its effectiveness further in Section~\ref{sec:eval}.

\begin{table}[ht!]
\caption{Extended variables and parameters of PLSCH-MTD.}
\label{tab:plschmtd}
\centering
\begin{tabular}{|c||c|c|l|} 
 \hline
 \textbf{Type} & \textbf{Symbol} & \textbf{Set} & \textbf{Definition} \\ [0.5ex] 
 \hline\hline
 Base & $c, e$ & $C$  & A configuration \\ \hline
 \multirow{2}{*}{Constant} & $\kappa$ & $Z^{*}$  &  Distance threshold \\ \cline{2-4}
 & $\alpha_{ce}$ & $Z^{*}$  & Indicates if $e$ can be configured after $c$  \\ \hline 
  Variable & $a_{ct}$ & $Z^{*}$  &  Decides if configuration $c$ set at $t$\\ \hline
 \end{tabular}
\end{table}

In PLSCH-MTD, we calculate the eligibility of each combination of potential configurations $c \in C$ in advance, considering a threshold distance $\kappa$ given as input. Two configurations can be set consecutively only if there is a sufficient amount of changes in-between s.t., $|c - e| > \kappa$, which is represented as $\alpha_{ce} = 1$. Table~\ref{tab:plschmtd} shows the new parameters and variables introduced with PLSCH-MTD. Accordingly, constraint~\ref{const:eligible} ensures that any consecutive MTD actions involve two eligible configurations satisfying the given threshold distance.
\begin{align}
y_{mjt} + y_{mkf} - 1 \leq \sum_{c \in C}^{}{\sum_{e \in C}^{}{a_{ct}a_{ef}\alpha_{ce}}} \qquad & &   \notag \\
  \qquad   \forall m \in M, \forall j,k \in J_m, k = j + 1, \forall t,f \leq T & &  \label{const:eligible} 
\end{align}
$a_{ct}$ is a binary decision variable representing if the system is reconfigured with configuration $c$ at time instance $t$. 
The quadratic expression in constraint~\ref{const:eligible} is linearized by using McCormick envelopes~\cite{Mccormick1976} to solve the problem easily with state-of-the-art linear optimization tools. Constraint~\ref{const:conf_placement} ensures that a respective configuration is assigned at $t$ if there is a defensive action taken s.t. $x_t = 1$.
\begin{align}
& \sum_{c \in C}^{}{a_{ct}} \leq x_t & \forall t \leq T \label{const:conf_placement} 
\end{align}
Lastly, Constraint~\ref{const:conf_uniqueness} avoids the reuse of the same configuration for the given system duration $T$ to prevent an attacker to deduct a reconfiguration pattern.
\begin{align}
& \sum_{t=0}^{T}{a_{ct}} \leq 1 & \forall c \in C \label{const:conf_uniqueness} 
\end{align}

Note that PLSCH-MTD is an offline solution in which the defender develops a strategy \emph{in advance} against several potential attack scenarios. Therefore, an increasing variety of considered attack scenarios could offer better strategies against broader threats. 
Besides, it does not require attack detection but can still prevent ongoing attacks by changing the service configuration. It also forces an attacker to rediscover the system with a new configuration.  In this sense, it is also a complementary security solution to reactive security mechanisms such as intrusion detection and prevention systems. 

\section{Attack Scenarios} \label{sec:attacker}

Several authors of related work tackle single attack scenarios conducted via real security tools. However, they cannot provide an optimal MTD schedule against multiple potential attack patterns~\cite{Thompson2014, Thompson2016}. More theoretical related work does not reflect realistic attacks well since they only use probability distribution functions for attack generation~\cite{Jones2015, Jones2017}.
Moreover, data on actual attacks against MCSs is limited to public reports and white papers that partially include attack durations and lack details regarding a complete attack timeline~\cite{Pols2017}. Although we know rough estimations on the time required for detecting advanced persistent threats~\cite{fireeye2020, mandiant2021} and detailed technical analysis of some infamous cyber-attacks and malware~\cite{albright2011stuxnet,Zhang2020}, it is difficult to obtain the complete picture of specific attack paths and the duration of advanced attacks.

Accordingly, we model different attack types and scenarios considering the recent security incidents in MCSs. An attack scenario is the combination of several individual attack steps as modeled in Section~\ref{sec:plsch}. Those scenarios are then used to evaluate the defensive strategies that PLSCH-MTD provides.

\subsection{Time Characteristics of Individual Attacks} \label{sec:time}

We define three attack types in terms of their duration: long, medium-length, and short attacks. The length of an attack represents its time-to-success value in the PLSCH model. Moreover, we introduce a new variable, $\Lambda$, the \emph{attack scale}, to set the relative lengths of different attacks in proportion to a common design parameter. It is defined in a similar scale with the time horizon $T$ (see Section~\ref{sec:plsch}) for a consistent representation of time-related variables. Accordingly, the length of each attack is uniformly sampled from an interval proportional to $\Lambda$. The attack types are characterized as follows: 

\begin{itemize}[leftmargin=*]
  \item \textbf{Long attack:} It represents the longest phases of an attack scenario, e.g., reconnaissance, developing necessary tools, and executing relatively complicated attack steps. The length of long attacks is sampled from the range of $[0.1 \Lambda, 0.3\Lambda]$, s.t., it lasts 20\% of a scenario with 10\% deviation for $\Lambda = T$.  

  \item \textbf{Medium-length attack:} It represents a certain number of successive attack steps that require significant time, e.g., encrypting a large amount of data or doing lateral movement across different network components. A medium-length attack is sampled from the interval $[0.05 \Lambda, 0.15 \Lambda]$, s.t. it typically takes 10\% of a scenario with 5\% time deviation. 

  \item \textbf{Short attack:} It represents a combination of successive attack steps with short execution time, e.g., changing the configuration of a component, modifying log files, etc. Their length is sampled from the interval $[0.0025 \Lambda, 0.075 \Lambda]$ taking on average 5\% of an attack scenario.
\end{itemize}

\subsection{Composition of Attack Scenarios}

\begin{figure}[t!]
\center
\includegraphics[scale=0.4]{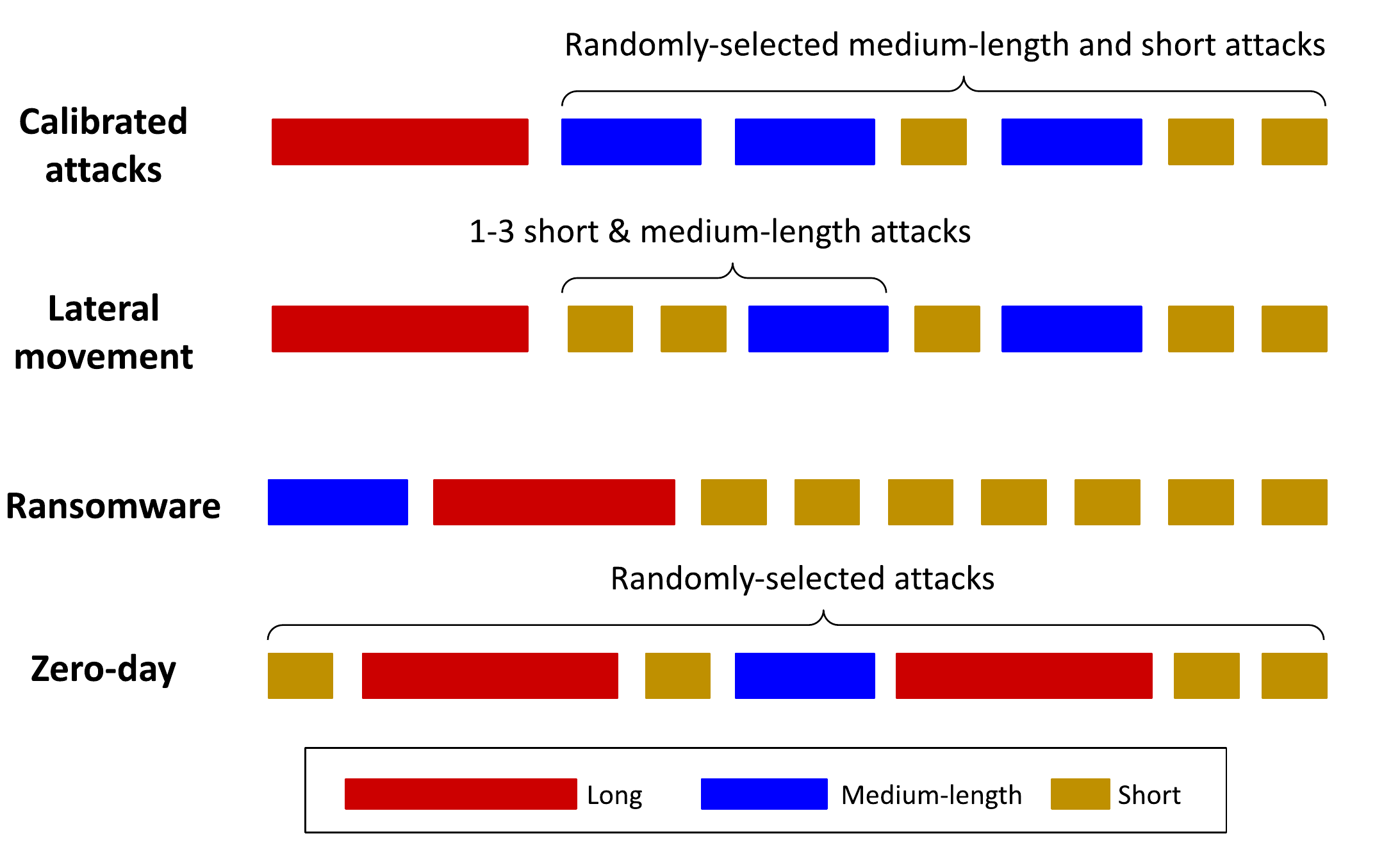}
\caption{Different attack scenarios.}
\label{fig:attack-scenarios}
\end{figure}

We in the following define four attack scenarios that reflect recent security incidents targeting critical networked systems~\cite{Chen2011, albright2011stuxnet, Wilkens2019, Olaimat2021}. They are composed of the attacks described above in dependence on different attacker goals as illustrated in Fig.~\ref{fig:attack-scenarios}. The duration of an attack scenario, i.e., the total lengths of its individual attack steps, is limited by the time horizon $T$ as it is also considered as the operational time of the system in the PLSCH model.

\begin{itemize}[leftmargin=*]
  \item \textbf{Calibrated attacks:} Calibrated attacks target specific components, technologies, and protocols in an MCS, e.g., although Stuxnet only damages a particular software that operates nuclear centrifuges~\cite{Chen2011}. Therefore, they require detailed system-specific knowledge and special exploits that induce long reconnaissance and development times. After acquiring access to the system, the attacker conducts a well-targeted sequence of attacks to potentially multiple components. Depending on the target, such attack steps can take different duration to reconnaissance and can be repeated several times~\cite{albright2011stuxnet}. Accordingly, we compose calibrated attack scenarios of (i) an initial \emph{long} attack and then (ii) randomly selected \emph{medium-length} and \emph{short} attacks as many as their total duration stays under $T$.

  \item \textbf{Lateral movement:} After gaining access to the system, an attacker can move laterally through the network to find critical services or sensitive data. While this still requires an initial reconnaissance time, the attacker should also discover further vulnerabilities to continue its lateral movement~\cite{Wilkens2019}, which imposes relatively shorter discovery campaigns. Meanwhile, gaining access to the other components potentially requires conducting more spontaneous attacks, e.g., acquiring credentials, patching legitimate software, etc. Accordingly, we compose lateral movement scenarios of (i) an initial \emph{long} attack for reconnaissance, (ii) several \emph{short} attacks for exploitation (between one to three attacks in our model), (iii) \emph{medium-length} discovery periods to move laterally, and (iv) repeating (ii) and (iii) steps through the movement until their total duration reaches to $T$.  

  \item \textbf{Ransomware:} Ransomware attacks spread a generic malware to encrypt files on the target systems and make them inaccessible. These attacks usually start with a phishing attempt, malvertising, or exploiting vulnerabilities in widely-used software~\cite{Olaimat2021}. Then, the attacker can wait a long time to discover the most sensitive data or cause the most damage to the target system at the right time. Lastly, it requires several operations for encrypting and copying the respective data. Accordingly, we compose these scenarios of (i) a \emph{medium-length} penetration time using one of the mentioned techniques, (ii) a \emph{long}(er) discovery and activation time, and (iii) \emph{short} operations for obtaining encrypted data as many as their total duration stays under $T$.

  \item \textbf{Zero-day:} Lastly, zero-day scenarios represent the threats that have not been encountered and thus not analyzed yet. They are composed of randomly-selected \emph{long}, \emph{medium-length}, and \emph{short} attacks with a total duration of $T$.
\end{itemize}

Note that $\Lambda$ only sets the proportion of time-to-success for individual attacks, and its value is not dependent on or limited by $T$. While higher $\Lambda$ values provide a higher number of shorter attacks, the opposite results in fewer but longer attacks. This enables us to specify attack scenarios for the desired time duration (depending on $T$) but still varying timing characteristics (depending on $\Lambda$) independently. 

For all scenarios, service reconfiguration within MTD actions helps to invalidate the attacker's knowledge about the system. For instance, service migrations can misorient attackers' movements in lateral movement scenarios. Similarly, against calibrated attacks, service reinitiations can recover the infected services and thus prevent their repetitive malicious behaviors. Finally, in ransomware scenarios, moving the backup data within database services, which is potentially discovered by the attacker, can prevent losing the sensitive data permanently and even disrupt the copying and the encryption processes. These changes also require establishing communication between reconfigured services and the rest of the system, i.e., rerouting data traffic over the network. In this sense, depending on the attack scenarios, the scope of the service reconfigurations can be specified for an MTD strategy. 

\section{Evaluation}\label{sec:eval}

PLSCH-MTD provides (i) feasible service configurations in terms of resource management and QoS for SOA-based MCSs via JSAR and (ii) optimal MTD schedules on the basis of these configurations to minimize the chance of a successful attack via PLSCH. 
Accordingly, in this section, we evaluate PLSCH-MTD by answering two main research questions: \\ \\
\textbf{RQ1:} \emph{How to find a sequence of effective service configurations that utilize the available configuration space efficiently and render the attacker's previous effort obsolete?} \\ 
\textbf{RQ2:} \emph{To which extent can an MTD schedule, which is restricted by a limited defender budget, protect the system against several potential attack scenarios?} \\

In the remainder of this section, we present our evaluation setup, the evaluation metrics, and the experimental results.

\subsection{Evaluation Setup}

We implemented our optimization models in CPLEX 12.7.0. All experiments were conducted on a server with 64-core Intel Xeon 2.10GHz CPU and 256GB RAM.
We generated random network topologies with $|V| = 20$ and an average connectivity of 1.7, and service overlays with $|D| = 15$ for each experiment as input to the JSAR.
The default time horizon $T$ and attack scale $\Lambda$ values are set to 60. Since we calculated the average number of attacks per scenario as 12 for the $T = \Lambda = 60$, and the time characteristics of attacks (see Section~\ref{sec:time}), we set the defender budget to 12 as well, i.e., sufficient budget to prevent all attacks in an ideal scenario.
The inter-configuration distance is set to 15\% in the joint model, PLSCH-MTD. All other parameter values are given within the respective experiment below.
Lastly, we perform 20 iterations per scenario to compute the average results with 95\% confidence interval.

\subsection{Evaluation Metrics} \label{sec:metrics}
We evaluate JSAR and PLSCH-MTD with different metrics:

\begin{itemize}[leftmargin=*]
  \item \textbf{Percentage of eligible configurations (PoEC):} This is the percentage of configurations that satisfy the minimum amount of required changes between two configurations, i.e., the inter-configuration distance. It indicates how flexible we can use the configuration space for successive MTD reconfigurations. 
  \item \textbf{Probability of retain (PoR):} It is the probability that a service instance or a data flow is not migrated after a reconfiguration. It represents whether an attacker can \textit{retain} access to the same service or the data traffic keeping its position, e.g., on the same node or link. 
  \item \textbf{Average attacker capture time (ACT):} It is measured by the ratio of the sum of all gaps between consecutive jobs across all machines in PLSCH to the total length of time horizons, i.e., $T*|M|$. The ACT represents the percentage of the total time that the attacker controls the system after a successful attack until the defender takes an MTD action. 
\end{itemize}

While the PoEC and the PoR measures \textit{the effective use of the configuration space} regarding \emph{RQ1}, the average attacker capture time~(ACT) measures \textit{the effectiveness of MTD scheduling} to examine \emph{RQ2}. 

\subsection{Experimental Results} \label{sec:results}
In this section, we present our numerical results.
For our experiments, we use several attack scenarios described in Section~\ref{sec:attacker}. Multiple instances of a particular scenario type vary due to the randomness in timing characteristics of each attack in a scenario, but still show similar scenario-specific patterns in terms of the order and length distribution of attacks. Then, PLSCH-MTD takes the generated scenarios for each type as input and provides an optimal MTD strategy against them. Alternatively, it is possible to defend against different scenario types at once, e.g., generating several instances per scenario simultaneously, which is referred to as \emph{mixed} scenarios in Section~\ref{sec:eval}. 
However, note that the length of individual attacks is selected consistently only within a respective scenario type. For instance, while a \emph{long} attack in calibrated attack scenarios can take months, it might be only days in a ransomware scenario. Therefore, it requires selecting a time scale for $\Lambda$ and $T$ that reasonably models all scenarios.  

\begin{figure}[h!]
\centering
\begin{subfigure}{.23\textwidth}
    \includegraphics[scale=0.28]{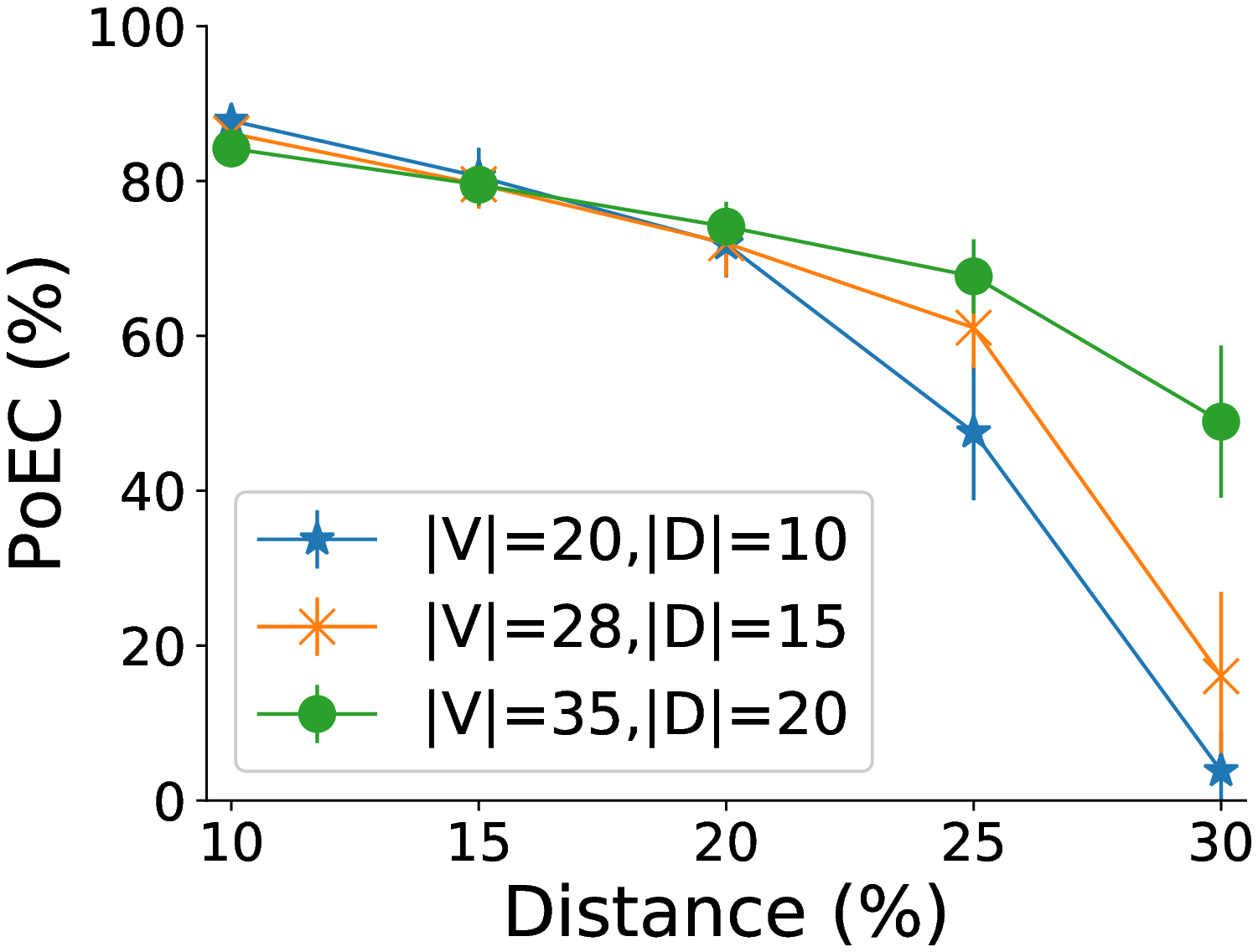}
    \caption{The impact on the PoEC}
    \label{fig:eligible_configs}
\end{subfigure}
~
\begin{subfigure}{.23\textwidth}
    \includegraphics[scale=0.28]{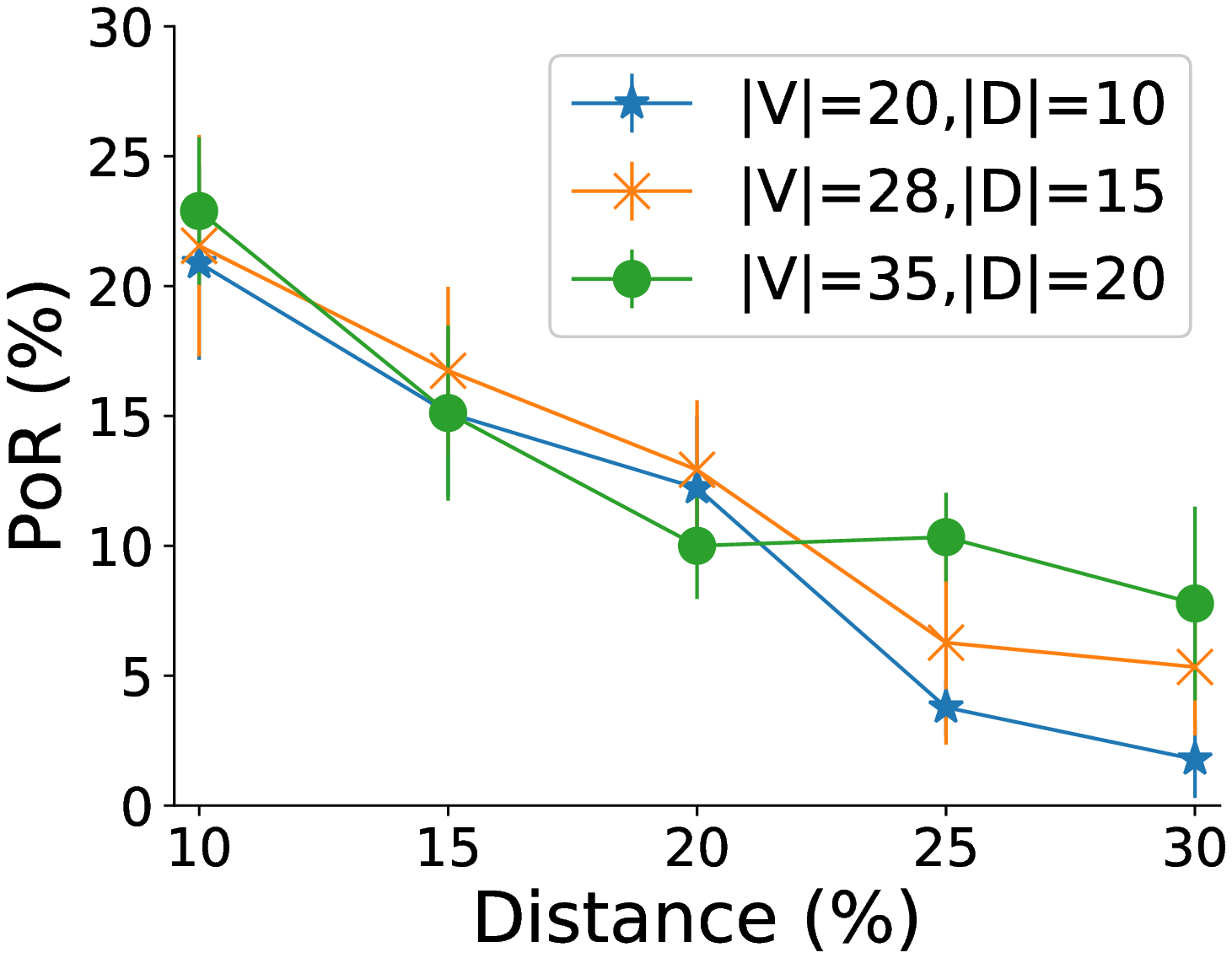}
    \caption{The impact on the PoR}
    \label{fig:attacker_reachability}
\end{subfigure}%
\caption{The impact of inter-configuration distance.}
\end{figure} 

\subsubsection{Effective use of the configuration space} 

We first evaluate the effective use of the potential configurations for MTD actions in terms of PoEC and PoR to answer the \emph{RQ1}. 

\textbf{The utilization of configuration space:} Fig.~\ref{fig:eligible_configs} shows PoEC for a changing percentage of the minimum inter-configuration distance~($\kappa$). The figure contains different graphs for the increasing size of network ($|V|$) and service overlays ($|D|$).
An increased minimum distance reduces the PoEC for each network size since it is getting harder to find configurations that are different enough, i.e., with a high inter-configuration distance due to stricter resource utilization. For small networks (blue, star), the percentage converges to nearly 0\% at 30\% minimum distance requirement. For a larger network (green, dot), in contrast, still around 60\% of the potential configurations can be used to reconfigure the service distribution and routing. 
The results in Fig.~\ref{fig:eligible_configs} indicate that the distance parameter is decisive on (i) having several potential configurations with fewer differences in between or (ii) fewer configurations with more substantial changes. On the one hand, the former enables a defender to utilize distinct configurations for a longer time frame and thus it is harder to detect a reconfiguration pattern for an attacker. On the other hand, a defender should use the same configurations repetitively in the latter scenario, which makes an MTD strategy easier detectable by attackers. 

\textbf{The impact of MTD reconfiguration on attackers:} Fig.~\ref{fig:attacker_reachability} shows the PoR for an increasing inter-configuration distance and different network sizes. 
While the 10\% distance threshold~($\kappa$) gives the attacker on the average a 20-25\% chance to access the same service or data that he attacked before the reconfiguration, it converges to nearly 0\% for small networks with 30\% minimum inter-configuration distance. 
For larger networks, the PoR is still as low as 10\% with a large confidence interval. The reason is, that although the solution space is larger, we do not select particular configurations, e.g., with the maximum distance, but arbitrarily select any two configurations that satisfy the minimum distance requirement. Although an arbitrary selection makes the next configuration less predictable for the attacker, the selection strategy can be adapted, e.g., selecting the configuration with the $k$th highest distance, to increase his reconnaissance effort. 
Consequently, a higher inter-configuration distance leads to further changes and enforces the attacker to rediscover the new configuration. However, it may also cause service interruptions. 

\subsubsection{Effectiveness of MTD scheduling}
We measure the impact of various parameters on the ACT to answer \emph{RQ2}, i.e., how protective an optimal MTD schedule is. 

\begin{figure}[t!]
\centering
\begin{subfigure}{.23\textwidth}
    \centering
    \includegraphics[scale=0.28]{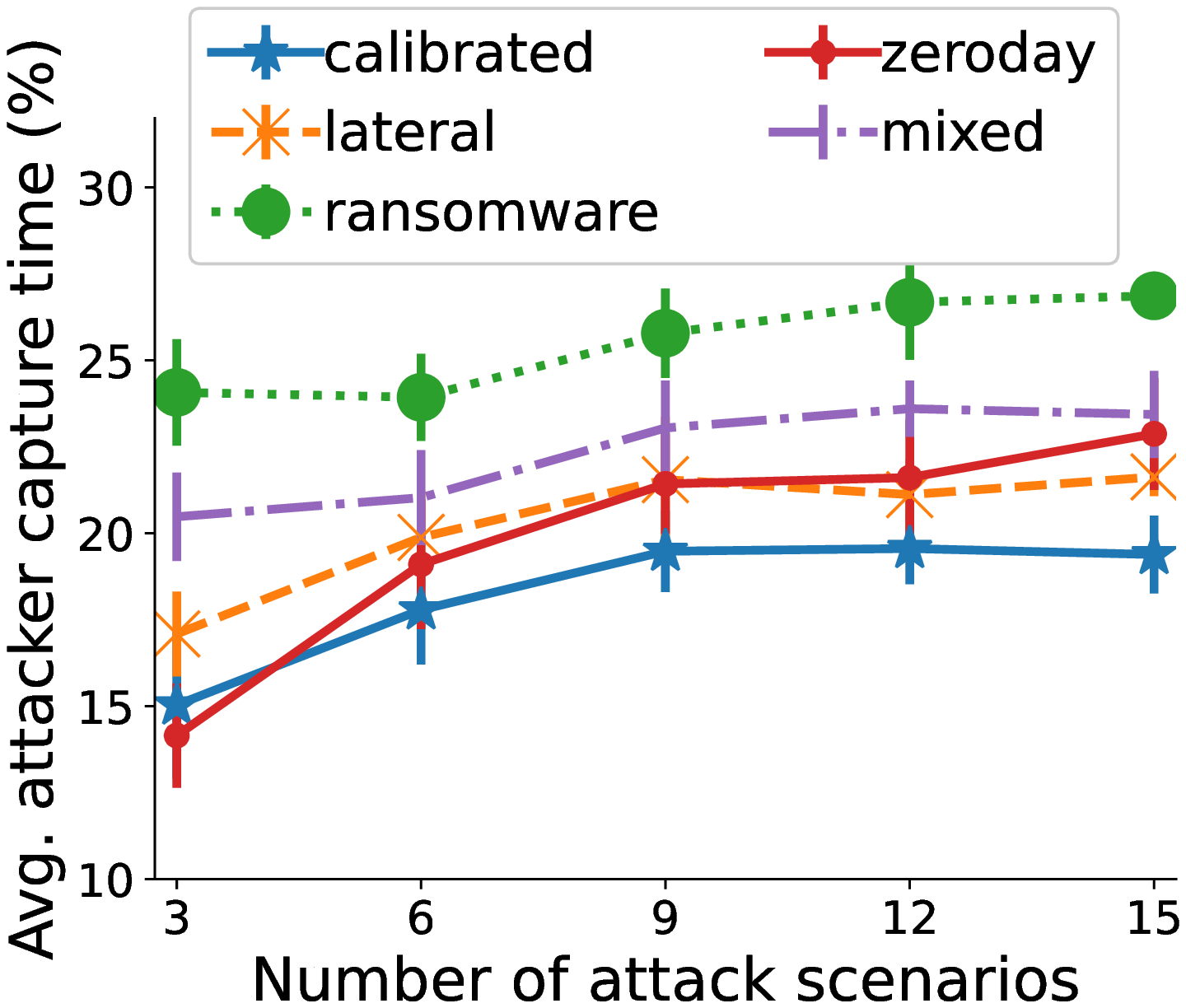}
\caption{Increasing number of scenarios}
    \label{fig:capture_machine}
\end{subfigure} 
~
\begin{subfigure}{.23\textwidth}
\includegraphics[scale=0.28]{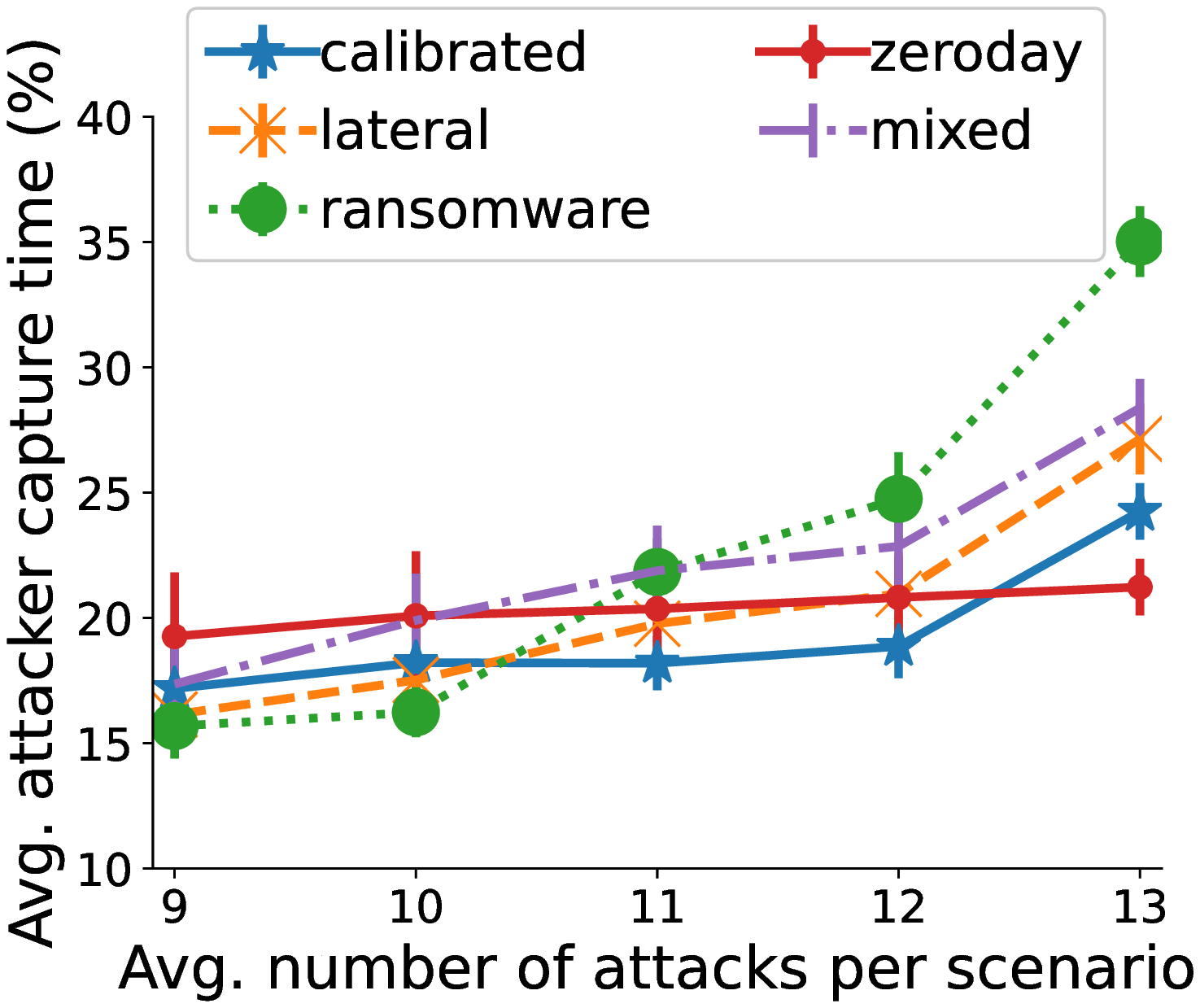}
\caption{Increasing number of attacks}
\label{fig:capture_attack}
\end{subfigure}
\caption{The impact of varying attack scenarios on the ACT.}
\end{figure}

\textbf{The impact of variety in attack scenarios:} We evaluate PLSCH-MTD for each type of attack scenario as well as for \emph{mixed} scenarios that include randomly-selected scenarios simultaneously. Fig.~\ref{fig:capture_machine} shows the impact of an increasing number of attack scenarios on the ACT.
Regardless of the scenario, we observe only a subtle increase from 1\% to 3\% in ACT. However, defending against multiple scenarios imposes a base challenge that results in 15-25\% ACT. The results indicate that although an MTD strategy remains protective against an increasing number of attack scenarios, it is still difficult to defend against even few concurrent scenarios.

The impact of the type of attack scenarios on ACT is more substantial than the impact of their quantity. In Fig.~\ref{fig:capture_machine}, defending against ransomware is the most challenging with 25\% ACT since it consists of several short attacks that can be accomplished. Other scenarios are similarly threatening with 15-18\% ACT. Therefore, the effectiveness of PLSCH-MTD is highly dependent on the actual attack scenario. 

To evaluate the impact of the number of attacks per scenario, we set $T = 60$ and $\Lambda \in [90, 50]$ (in a reversed order). Decreasing $\Lambda$ shortens the length of individual attacks and increases their number per scenario, which results in 9-13 attacks for the given range of $\Lambda$ values. Accordingly, Fig.~\ref{fig:capture_attack} shows the ACT measurements for increasing attacks per scenario. In the figure, the ACT does not significantly change for 9 to 12 attacks within each attack scenario as there is enough defender budget ($\beta=12$). This also affirms the results regarding the base challenge (15-25\%) of defending against multiple scenarios in Fig.~\ref{fig:capture_machine}. However, the attacker's success increases by 5-10\% for 13 attacks due to the insufficient defender budget.

\begin{figure}[h!]
    \centering
    \includegraphics[scale=0.32]{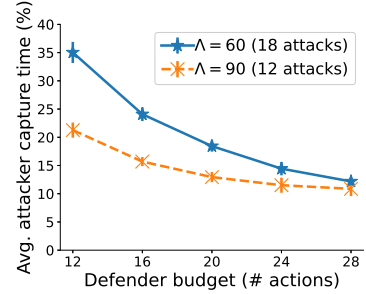}
    \caption{The impact of defender budget on the ACT.}
    \label{fig:budget}
\end{figure}

\textbf{The impact of defender budget:} Fig.~\ref{fig:budget} shows the impact of an increasing defender budget on the ACT for different numbers of attacks per scenario, i.e., for 18 and 12 attacks by setting $\Lambda=\{60, 90\}$ and $T = 90$. As seen in the figure, more budget strengthens the defender to hold control of the resources with a decreasing ACT regardless of attack counts. When the defender budget is less than the number of attacks per scenario, i.e., for $\beta = {12-16}$ and $\Lambda = 60$ (solid, blue line), we can observe that a gradual increase in the budget decreases the ACT from 35\% to 20\%. 

Theoretically, any defender budget $\beta \geq T$ should guarantee a complete defender occupation. This enables \textit{moving} the system at every possible time instance $t \leq T$ (which is infeasible in practice due to its high overhead) and thus leaves the attacker no chance to accomplish an attack. 
However, as shown in Fig.~\ref{fig:budget}, it is not possible to obtain that level of protection quickly with a linear increase in the budget after the ACT has converged to 10-12\% due to the challenges in defending against multiple attack scenarios. 

Note that a single successful attack step may not give the attacker total control over the system as assumed in the PLADD game, but it requires several attack steps to be accomplished. In this sense, our ACT measurements represent \emph{the worst case} that each attack is equally effective. As a result, PLSCH-MTD can still achieve protection of up to 90\% of the system operational time with a defensive budget $\beta \ll T$.

\section{Conclusion} \label{sec:conclusion}

Service-oriented architecture~(SOA) enables the flexible design of mission-critical systems~(MCSs) by dynamically distributing virtual services and establishing their inter-communication. This flexibility can also be utilized to implement moving target defense~(MTD) strategies for the security of MCSs. By reconfiguring the critical services and data traffic periodically within MTD strategies, it is possible to protect MCSs against advanced cyber-attacks. In this work, we propose an optimization framework~(PLSCH-MTD) by combining a joint service allocation and routing model~(JSAR) with an attack-defender game~(PLSCH) to find effective MTD strategies for SOA-based MCSs. While PLSCH provides an optimal schedule of subsequent MTD actions against potential threats, JSAR generates feasible service configurations for each action. Furthermore, we model several attack scenarios inspired by the security incidents in MCSs to evaluate PLSCH-MTD. The experiments reveal that the PLSCH-MTD can utilize the service configuration space efficiently to force attackers to rediscover the system. Moreover, it can protect an MCS for up to 90\% of its operational time. 
\newpage
\balance
\bibliographystyle{ieeetr}
\bibliography{references}
\end{document}